\documentclass[12pt]{article}
\usepackage{times}
\usepackage{geometry}
\usepackage[utf8]{inputenc}
\usepackage{enumitem,amssymb}
\usepackage{ragged2e}
\usepackage{graphicx}
\usepackage{comment}
\usepackage{multicol}
\usepackage[usenames]{xcolor} %used for font color              
\definecolor{xlinkcolor}{cmyk}{1,1,0,0}
\usepackage{url}
\usepackage[
 colorlinks=true,    % false: boxed links; true: colored links 
 linkcolor=xlinkcolor,     % color of internal links            
 citecolor=xlinkcolor,     % color of links to bibliography
 filecolor=xlinkcolor,  % color of file links 
 urlcolor=xlinkcolor,      % color of external link
 final=true
]{hyperref}
\usepackage[super,sort&compress]{natbib}
\usepackage{enumitem}
\setenumerate{itemsep=0mm}

\newlist{thematic}{itemize}{8}
\setlist[thematic]{label=$\square$}
\usepackage{pifont}

\usepackage{macros}
\usepackage{aas_macros}

\begin{document}
\begin{raggedright} 
% part of template, but does not look good
\huge
Astro2020 Science White Paper \hfill ~ \linebreak
Dark Matter Science in the Era of LSST \hfill ~ \linebreak
\end{raggedright}
\normalsize

\noindent \textbf{Thematic Areas:} \hspace*{58pt} $\square$ Planetary Systems \hspace*{12pt} $\square$ Star and Planet Formation  \hfill ~\linebreak
$\blacksquare$ Formation and Evolution of Compact Objects \hspace*{31pt} $\blacksquare$ Cosmology and Fundamental Physics \hfill ~ \linebreak
  $\square$  Stars and Stellar Evolution \hspace*{1pt} $\square$ Resolved Stellar Populations and their Environments \hspace*{40pt} \hfill ~\linebreak
  $\square$    Galaxy Evolution   \hspace*{45pt} $\square$             Multi-Messenger Astronomy and Astrophysics \hspace*{65pt} \hfill ~ \linebreak

% Author list file generated with: mkauthlist.py UNKNOWN 
% python code/mkauthlist.py -f -s -j emulateapj -a data/order.csv data/astro2020_endorsers_trimmed_merged.csv tmp.tex 

\def\altaffilmark#1{\textsuperscript{#1}}
\def\affil#1{\noindent #1 \\}

\noindent \textbf{Principal Authors:} 
Keith Bechtol\altaffilmark{1}, \href{mailto:kbechtol@wisc.edu}{kbechtol@wisc.edu} (UW Madison)\\
Alex~Drlica-Wagner\altaffilmark{2,3,4}, \href{mailto:kadrlica@fnal.gov}{kadrlica@fnal.gov} (Fermilab/KICP/UChicago)

%\noindent Name:	
% \linebreak						
%Institution:  
% \linebreak
%Email: 
% \linebreak
%Phone:  

\noindent {\bf Co-authors (affiliations after text):}
\begin{raggedright}
\small
Kevork N.\ Abazajian\altaffilmark{5},
Muntazir Abidi\altaffilmark{6},
Susmita~Adhikari\altaffilmark{7},
Yacine~Ali-Ha\"imoud\altaffilmark{8},
James~Annis\altaffilmark{2},
Behzad Ansarinejad\altaffilmark{9},
Robert~Armstrong\altaffilmark{10},
Jacobo Asorey\altaffilmark{11},
Carlo Baccigalupi\altaffilmark{12,13,14},
Arka~Banerjee\altaffilmark{7,15}
% ...

\end{raggedright}

\noindent \textbf{Abstract:}
Astrophysical observations currently provide the only robust, empirical measurements of dark matter. 
%Astrophysical probes will continue to guide other experimental efforts into the next decade, 
%In the coming decade, astrophysical probes of dark matter will explore parameter space beyond the current sensitivity of the high-energy physics program and will complement future experimental searches.
%Future observations with the Large Synoptic Survey Telescope (LSST) will provide necessary guidance for the experimental dark matter program. 
In the coming decade, astrophysical observations will guide other experimental efforts, while simultaneously probing unique regions of dark matter parameter space.
This white paper summarizes astrophysical observations that can constrain the fundamental physics of dark matter in the era of LSST. 
We describe how astrophysical observations will inform our understanding of the fundamental properties of dark matter, such as particle mass, self-interaction strength, non-gravitational interactions with the Standard Model, and compact object abundances. Additionally, we highlight theoretical work and experimental/observational facilities that will complement LSST to strengthen our understanding of the fundamental characteristics of dark matter.
%More information on the LSST dark matter effort can be found at \href{https://lsstdarkmatter.github.io/}{https://lsstdarkmatter.github.io/}.

\pagebreak

% Insert your white paper text here (max 5 pages inc. figs).

\vspace{-1em} \subsection*{Summary}

More than 85 years after its astrophysical discovery, the fundamental nature of dark matter remains one of the foremost open questions in science.
% physics and astronomy
Over the last several decades, an extensive experimental program has sought to determine the cosmological origin, constituents, and interaction mechanisms of dark matter. 
%While the existing experimental program has largely focused on weakly-interacting massive particles, there is strong theoretical motivation to explore a broader set of dark matter candidates.
%As the high-energy physics program expands to ``search for dark matter along every feasible avenue'' \citep{P5Report}, it is essential to keep in mind that the only direct, empirical measurements of dark matter properties to date come from astrophysical and cosmological observations.
%More than 85 years after its astrophysical discovery, the fundamental nature of dark matter remains one of the foremost open questions in physics and astronomy.
%Over the last several decades, an extensive experimental program has sought to determine the cosmological origin, fundamental constituents, and interaction mechanisms of dark matter.
To date, the only direct, positive empirical measurements of dark matter come from astrophysical observations.
%Discovering the fundamental nature of dark matter will draw upon the tools of particle physics, cosmology, stellar astrophysics, and galaxy evolution.
Discovering the fundamental nature of dark matter will necessarily draw upon the tools particle physics, cosmology, and astronomy.

LSST will provide a unique and impressive platform to study dark sector physics in the 2020s.
Originally envisioned as the ``Dark Matter Telescope'' \citep{Tyson:2001}, LSST will enable precision tests of the \LCDM model and elucidate the connection between luminous galaxies and the cosmic web of dark matter. 
Cosmology has consistently shown that it is impossible to separate the \emph{macroscopic distribution} of dark matter from the \emph{microscopic physics} governing dark matter.
In fact, some microscopic characteristics of dark matter are \emph{only accessible} via astrophysics.
Studies of dark matter, dark energy, massive neutrinos, and galaxy evolution are \emph{extremely complementary} from both a technical and scientific standpoint. 
A robust dark matter program leveraging LSST data has the ability to test a broad range of well-motivated theoretical models of dark matter including self-interacting dark matter, warm dark matter, dark matter-baryon scattering, ultra-light dark matter, axion-like particles, and primordial black holes. 

LSST will enable studies of Milky Way satellite galaxies, stellar streams, and strong lens systems to detect and characterize the smallest dark matter halos, thereby probing the minimum mass of ultra-light dark matter and thermal warm dark matter.
Precise measurements of the density and shapes of dark matter halos in dwarf galaxies and galaxy clusters will be sensitive to dark matter self-interactions probing hidden sector and dark photon models.
Microlensing measurements will directly probe primordial black holes and the compact object fraction of dark matter at the sub-percent level over a wide range of masses.
Precise measurements of stellar populations will be sensitive to anomalous energy loss mechanisms and will constrain the coupling of axion-like particles to photons and electrons.
Measurements of large-scale structure will spatially resolve the influence of both dark matter and dark energy, enabling searches for correlations between the two known components of the dark sector.
In addition, complementarity between astrophysical, direct detection, and other indirect searches for dark matter will help constrain dark matter-baryon scattering, dark matter self-annihilation, and dark matter decay. 

%Studies of dark matter with LSST will provide critical information about the fundamental nature of dark matter over the next decade at a low cost by leveraging a soon-to-exist facility.
%By leveraging a soon to-exist-facility, a small program with LSST will provide critical information about the fundamental nature of dark matter over the next decade at a low cost. 
%The study of dark matter with LSST presents a small experimental program that is guaranteed to provide critical information about the fundamental nature of dark matter over the next decade.
%LSST will rapidly produce high-impact science on the nature of fundamental dark matter by exploiting a soon-to-exist facility. 
Astrophysical dark matter studies will explore parameter space beyond the current sensitivity of the high-energy physics program and will complement other experimental searches.
This has been recognized in Astro 2010 \citep{Astro2010}, during the Snowmass Cosmic Frontier planning process \citep[][]{1310.8642, 1310.5662, 1305.1605}, in the P5 Report \citep[]{P5Report}, and in a series of recent Cosmic Visions reports \citep[][]{1604.07626,1802.07216}, including the ``New Ideas in Dark Matter 2017:\ Community Report'' \citep{Battaglieri:2017aum}.
%Astrophysical probes provide the only constraints on the minimum and maximum mass scale of dark matter, and 
%Astrophysical observations will likely continue to guide other experimental efforts.
%the experimental particle physics program for years to come.
In the 2020s, the impact of the LSST dark matter program will be enhanced by access to wide-field massively multiplexed spectroscopy on medium- to large-aperture telescopes ($\roughly 8$--$10$-meter class), deep spectroscopy on giant segmented mirror telescopes ($\roughly 30$-m class), together with high-resolution optical and radio imaging.
%with relatively smaller fields of view
Further theoretical work is also needed to interpret those observations in terms of particle models, to combine results from multiple observational methods, and to develop novel probes of dark matter.

This whitepaper is a summary of Drlica-Wagner et al. (2019) \citep{drlica-wagner_2019_lsst_dark_matter}.

\begin{table}[ht]
\footnotesize
\begin{center}
\begin{tabular}{l c c c}
\hline 
Model & Probe & Parameter & Value \\
\hline 
\hline
Warm Dark Matter  & Halo Mass & Particle Mass & $m \sim 18 \keV$ \\
Self-Interacting Dark Matter & Halo Profile & Cross Section & $\sigmam \sim 0.1\text{--}10\cm^2/\g$ \\
Baryon-Scattering Dark Matter & Halo Mass & Cross Section & $\sigma \sim 10^{-30} \cm^2$ \\
Axion-Like Particles & Energy Loss & Coupling Strength & $g_{\phi e} \sim 10^{-13} $ \\
Fuzzy Dark Matter & Halo Mass & Particle Mass & $m \sim 10^{-20} \eV$  \\
Primordial Black Holes  & Compact Objects & Object Mass & $M > 10^{-4} \Msun$ \\
WIMPs & Indirect Detection & Cross Section & $\sigmav \sim 10^{-27} \cm^3/\second$ \\
Light Relics & Large-Scale Structure & Relativistic Species & $N_{\rm eff} \sim 0.1$ \\[+0.5em]
\hline
\end{tabular}
\end{center}
\vspace{-1em}
\caption{\label{tab:models} Probes of fundamental dark matter physics in the LSST era, organized by dark matter model and associated observables. Sensitivity forecasts appear in the rightmost column.}
%Probes of fundamental dark matter physics with LSST. The four columns indicate classes of dark matter models, primary observational probe, corresponding dark matter parameters, and the estimated senstivity of LSST.}
%Sensitivity forecasts of Probes of fundamental dark matter physics in the LSST era. 

%Classes of dark matter models are listed in Column 1, and the primary observational probe that is sensitive to each model is listed in Column 2. The corresponding dark matter parameters are listed in Column 3, and estimates of LSST's senstivity to each parameter are listed in Column 4.}
\end{table}

\vspace{-1em} \subsection*{Dark Matter Models} \vspace{-0.5em}

%Astrophysical observations use gravity to directly probe dark matter. 
Astrophysical observations probe the physics of dark matter through its
impact on structure formation throughout cosmic history.
On large scales, current observational data are well described by a simple model of stable, non-relativistic, collisionless, cold dark matter (CDM).
However, many viable theoretical models of dark matter predict deviations from CDM that are testable with current and future observations.
Fundamental properties of dark matter---e.g., particle mass, self-interaction cross section, coupling to the Standard Model, and time evolution---can imprint themselves on the macroscopic distribution of dark matter in a detectable manner. With supporting theoretical efforts and follow-up observations, LSST will be sensitive to several distinct classes of dark matter models, including particle dark matter, field dark matter, and compact objects (\tabref{models}).

\noindent \textbf{Particle Dark Matter:} LSST, in combination with other observations, will be able to probe microscopic characteristics of particle dark matter such as self-interaction cross section, particle mass, baryon-scattering cross section, self-annihilation rate, and decay rate. These measurements will complement and guide collider, direct, and indirect detection efforts to study particle dark matter.

%Minimum halo mass, halo profiles, compact object abundance, anomalous energy loss mechanisms, and large-scale structure.

\noindent \textbf{Wave-like Dark Matter:} Axion-like particles and other (ultra-)light dark matter candidates are a natural alternative to conventional particle dark matter. LSST will be uniquely sensitive to the minimum mass of ultra-light dark matter and to couplings between axion-like particles and the Standard Model.

\noindent \textbf{Compact Objects:} Compact object dark matter is fundamentally different from particle models; primordial black holes cannot be studied in an accelerator and can only be detected through their gravitational force. 
Primordial black holes (PBHs) formed directly from the primordial density fluctuations could make up some fraction of the dark matter, and a measurement of their abundance would directly constrain the amplitude of density fluctuations and provide unique insights into physics at ultra-high energies.

\vspace{-1em} \subsection*{Dark Matter Probes} \vspace{-0.5em}

\noindent {\bf Minimum Halo Mass:}
The standard cosmological model predicts a nearly scale-invariant mass spectrum of dark matter halos down to Earth-mass scales (or below), e.g., in WIMP and non-thermal axion models \citep{Green:2003un,2005Natur.433..389D,1412.5930}.
%A defining prediction of the standard cosmological model with cold dark matter (CDM) is the gathering of dark matter into gravitationally bound halos having a nearly scale-invariant mass spectrum on physical scales ranging from galaxy clusters to planet-scale masses.
%The cold, collisionless model of dark matter predicts that dark matter halos should exist down to Earth-mass scales (or below) in WIMP and non-thermal axion models \citep{Green:2003un,2005Natur.433..389D,1412.5930}.
Modifications to the cold, collisionless dark matter paradigm can suppress the formation of dark matter halos on these small scales.
Current observations provide a robust measurement of the dark matter halo mass spectrum for halos with mass $> 10^{10}\Msun$, and the smallest known galaxies provide an existence proof for halos of mass $\roughly 10^8 \Msun - 10^9 \Msun$ \citep{2017MNRAS.467.2019R,behroozi2018,Jethwa:2018,Kim:2017iwr,Nadler:2018,1807.07093}. 
LSST will expand the census of ultra-faint satellite galaxies orbiting the Milky Way and enable statistical searches for extremely low-luminosity and low-surface brightness galaxies throughout the Local Volume.
By measuring the galaxy luminosity function at the extreme low-mass threshold of galaxy formation, LSST will test the abundance of dark matter halos at $\sim10^8 \Msun$.

LSST will probe dark matter halos below the threshold of galaxy formation with stellar streams and strongly lensed systems.
%LSST will enable searches for completely dark halos using purely gravitational observational signatures, e.g., stellar stream gaps and strong gravitational lensing anomalies.
%subhalos purely through their gravitational signatures
Galactic dark matter subhalos with masses as small as $10^5$--$10^6 \Msun$ passing a stellar stream are capable of producing detectable gaps in the stellar density \citep[][]{erkal2016,bovy:2017}.
%Deep and precise LSST photometry is expected to increase the contrast between streams and the contaminating Milky Way field stars, dramatically increasing our ability to detect density variations and thus leading to the identification of less prominent gaps created by low-mass perturbers
By identifying additional stellar streams and increasing the density contrast of known streams against the smooth Milky Way halo, LSST will shift analysis from individual gaps into the regime of subhalo population statistics and (in)consistency with cold dark matter predictions.
Importantly, LSST will allow studies of streams farther from the center of the Galaxy for which confounding baryonic effects are lessened.
%LSST will mitigate both of these issues by examining streams farther from the center of the Galaxy where these effects are lessened.
Meanwhile, strong gravitational lensing can be used to measure the abundance and masses of subhalos in massive galaxies and small isolated halos along the line of sight at cosmological distances, independent of their baryon content.
LSST will increase the number of lensed systems from the current sample of hundreds to an expected sample of thousands of lensed quasars \citep{O+M10} and tens of thousands of lensed galaxies \citep{Collett2015}.
%Through analysis of flux ratio anomalies, gravitational imaging, and measuring the power spectrum of 

\noindent {\bf Halo Profiles:}
Measurements of the radial density profiles and shapes of dark matter halos are sensitive to the microphysics governing non-gravitational dark matter self-interactions, which could produce flat density cores \citep{Spergel:1999mh} and more spherical halo shapes \citep{Peter:2013}.
%Dwarf galaxies, galaxy clusters, merging clusters.
Through galaxy-galaxy weak lensing, LSST will be able to distinguish cored versus cuspy NFW density profiles for a sample of low-redshift dwarf galaxies with masses $M_\text{halo} = 3\times10^9\,h^{-1}\Msun$.
Studies of the density profiles of massive galaxy clusters, as well as systems of merging galaxy clusters, will constrain the scattering cross section at the level $\sigmam \sim 0.1-1 \cmg$.
Measuring halo profiles over a range of mass scales will provide sensitivity to dark matter scattering with non-trivial velocity dependence.
%Due to the possibility that dark matter scattering has a non-trivial velocity dependence, it is important to probe halo profiles over a wide range of mass scales.
%The standard CDM model predicts that dark matter halos should be “cuspy, i.e. with inner densityprofiles asymptoting to high central densities. If dark matter is able to interact through scattering or the exchange of some light mediator, then the density of halos could instead flatten out to produce dark matter “cores”.

\noindent {\bf Compact Objects:} 
LSST has the ability to directly detect signals of compact halo objects through precise, short- ($\roughly 30 \second$) and long-duration ($\roughly1 \unit{yr}$) observations of classical and parallactic microlensing\citep{1509.04899}.
If scheduled optimally, LSST could extend PBH sensitivity to $\roughly0.03\%$ of the dark matter fraction for masses $\gtrsim 10^{-1} \Msun$.
By supplementing the LSST survey with astrometric microlensing observations, it will be possible to break lensing mass-geometry degeneracies and make precise measurements of individual black hole masses. Thus, if PBHs make up a significant fraction of dark matter, LSST will effectively measure their ``particle'' properties and provide insight into the fundamental physics of the early universe.

\noindent {\bf Anomalous Energy Loss:}
Observations of stars provide a mechanism to probe temperatures, particle densities, and time scales that are inaccessible to laboratory experiments. Since conventional astrophysics allows us to quantitatively model the evolution of stars, detailed study of stellar populations can provide a powerful technique to probe new physics. In particular, if new light particles exist and are coupled to Standard Model fields, their emission would provide an additional channel for stellar energy loss. 
LSST will greatly improve our understanding of stellar evolution by providing unprecedented photometry, astrometry, and temporal sampling for a large sample of faint stars. In particular, measurements of the white dwarf luminosity function, giant branch stars, and core-collapse supernovae will provide sensitivity to the axion-electron coupling.

\noindent {\bf Large-Scale Structure:} LSST will produce the largest and most detailed map of the distribution of matter and the growth of cosmic structure over the past 10 Gyr.
The large-scale clustering of matter and luminous tracers in the late-time universe is sensitive to the total amount of dark matter, the fraction of dark matter in light relics that behave as radiation at early times, and fundamental couplings between dark matter and dark energy.
Measurements of large-scale structure with LSST will enhance constraints on massive neutrinos and other light relics from the early universe that could compose a fraction of the dark matter.
Additionally, LSST will use supernovae and $3\times2$pt statistics of galaxy clustering and weak lensing to measure dark energy in independent patches across the sky, allowing for spatial cross correlation between dark matter and dark energy \citep{0902.2590}.

\begin{figure}[t]
\centering
\includegraphics[width=0.53\columnwidth]{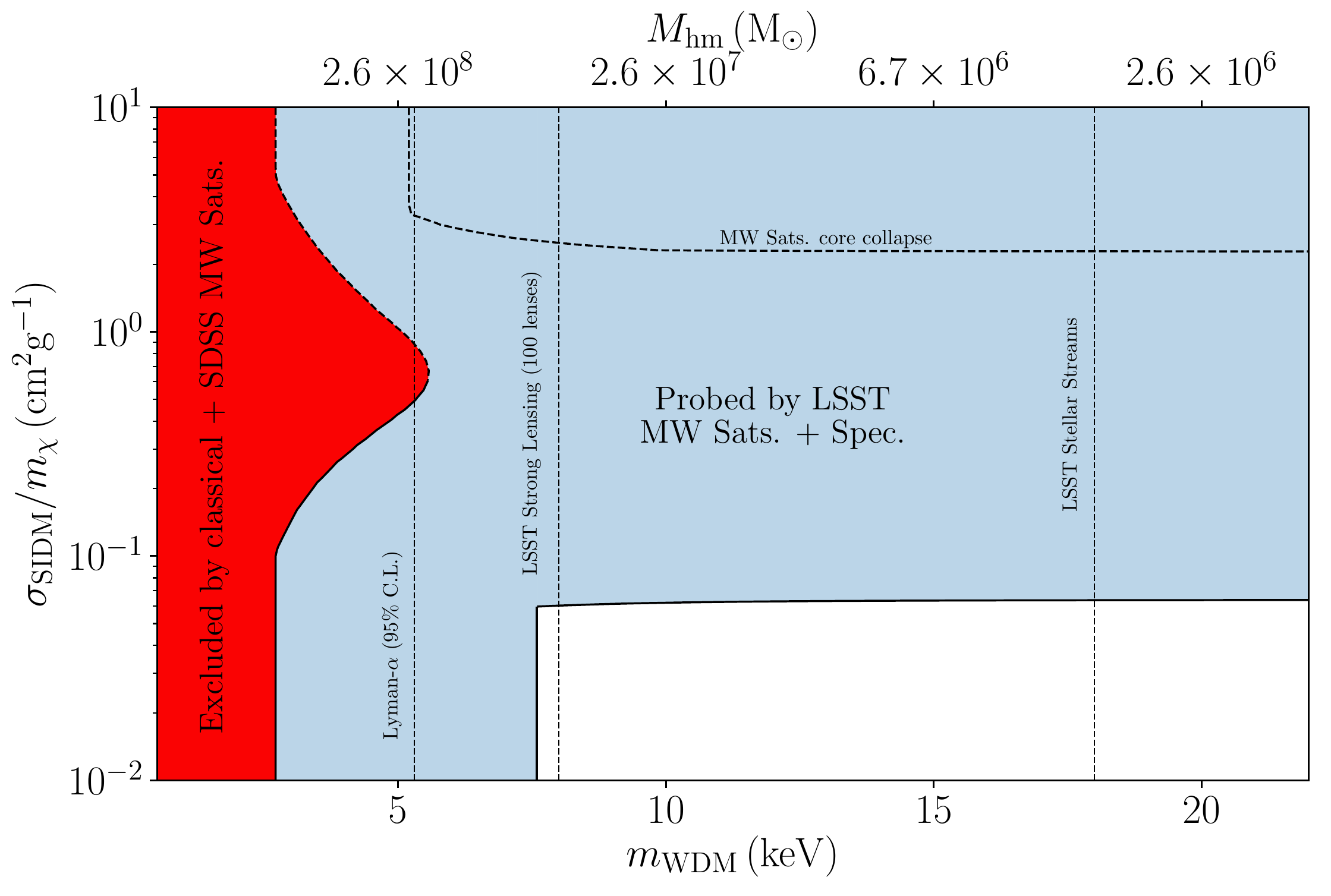}
\includegraphics[width=0.46\columnwidth]{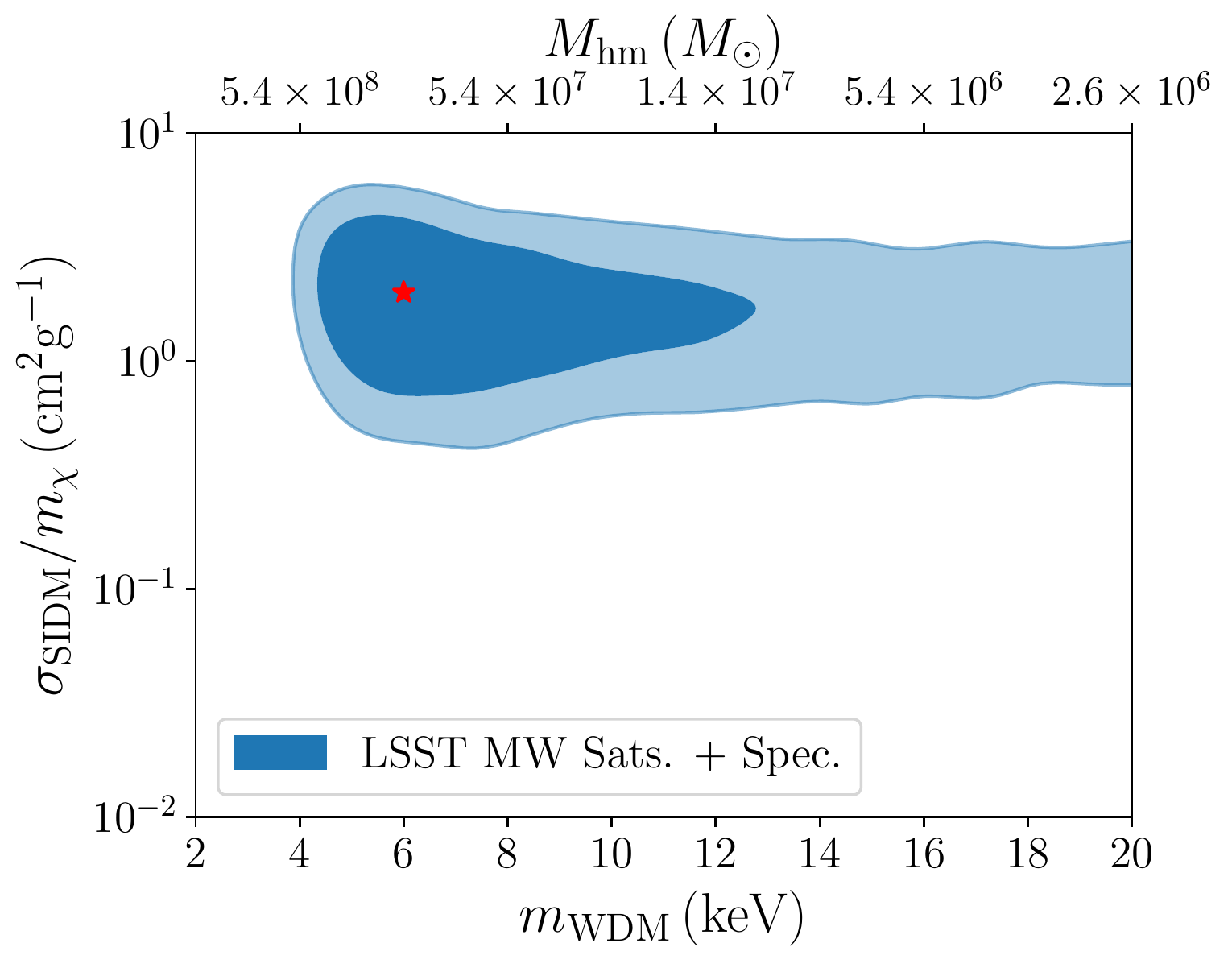}
\caption{\emph{Left}: Projected joint sensitivity to WDM particle mass and SIDM cross section from LSST observations of dark matter substructure. 
\emph{Right}: Example of a measurement of particle properties for a dark matter model with a self-interaction cross section and matter power spectrum cut-off just beyond current constraints ($\sigmam = 2 \cmg$ and $\mWDM = 6\keV$, indicated by the red star) \citep{drlica-wagner_2019_lsst_dark_matter}. Complementary observations can break degeneracies among dark matter models that have the same approximate behavior on small scales but differ in detail.}
\end{figure}

\vspace{-1em} \subsection*{Complementarity} \vspace{-0.5em}

%While LSST is the discovery engine, many other complementary observations are required to realize the astrophysical dark matter program.
LSST will enable complementary studies of dark matter with spectroscopy, high-resolution imaging, indirect detection experiments, and direct detection experiments.
While LSST can substantially improve our understanding of dark matter in isolation, the combination of experiments is essential to confirm future discoveries and provide a holistic picture of dark matter physics.
%the combination of experiments is essential to provide a holistic picture of dark matter physics.

\noindent {\bf Spectroscopy:}
Wide field-of-view, massively multiplexed spectroscopy on 8--10-meter-class telescopes as well as deep spectroscopy with 30-meter-class telescopes will complement studies of minimum halo mass and halo profiles.
% smaller field-of-view

\noindent {\bf High-Resolution Imaging:} High-resolution follow-up imaging at the milliarcsecond-scale from space and with ground-based adaptive optics are needed to maximize strong lensing, microlensing, and galaxy cluster studies with LSST.

\noindent {\bf Indirect Detection:} By precisely mapping the distribution of dark matter on Galactic and extragalactic scales, LSST will enable more sensitive searches for energetic particles created by dark matter annihilation and/or decay, e.g., using gamma-ray or neutrino telescopes \citep{Charles:2016,Albert:2017,1404.5503}.
LSST will also provide sensitivity to axion-like particles via monitoring extreme events in the transient sky \citep{2017PhRvL.118a1103M}.

%Leading constraints on the dark matter annihilation cross section come from gamma-ray analysis of Milky Way satellite galaxies.
%originating from the dark sector.
% KB: What is meant by "extreme events". Maybe a separate sentence?
%and tracking extreme events 

\noindent {\bf Direct Detection:} LSST will complement direct detection experiments by improving measurements of the local phase-space density of dark matter using precision astrometry of Milky Way stars.
For dark matter-baryon scattering, small-scale structure measurements with LSST can probe dark matter masses and cross sections outside the range accessible to direct detection experiments.

%\vspace{-1em} \subsection*{Outlook: Discovery Potential} \vspace{-0.5em}

\begin{figure}[t]
\centering
\includegraphics[width=0.49\textwidth]{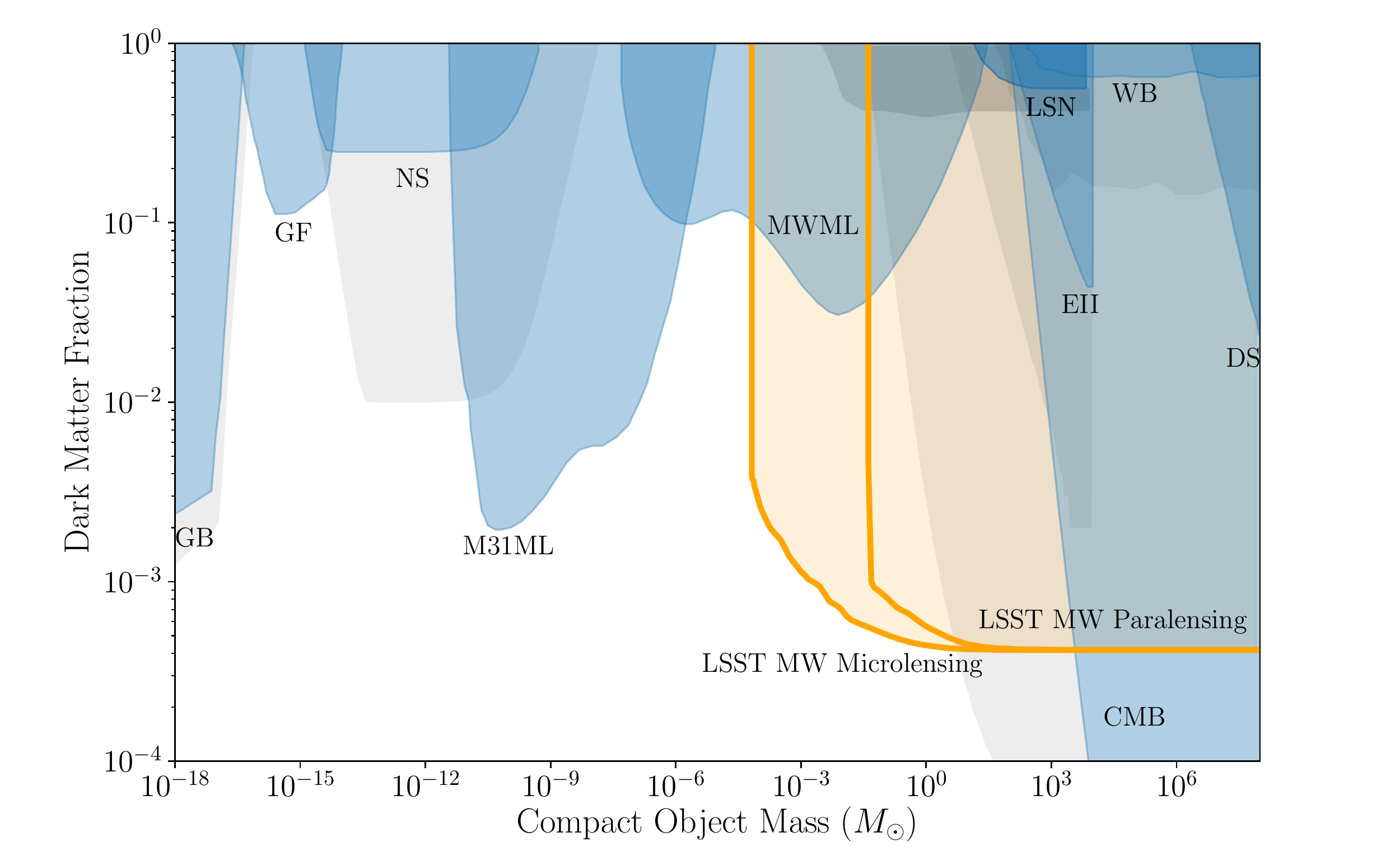}
\includegraphics[width=0.49\columnwidth]{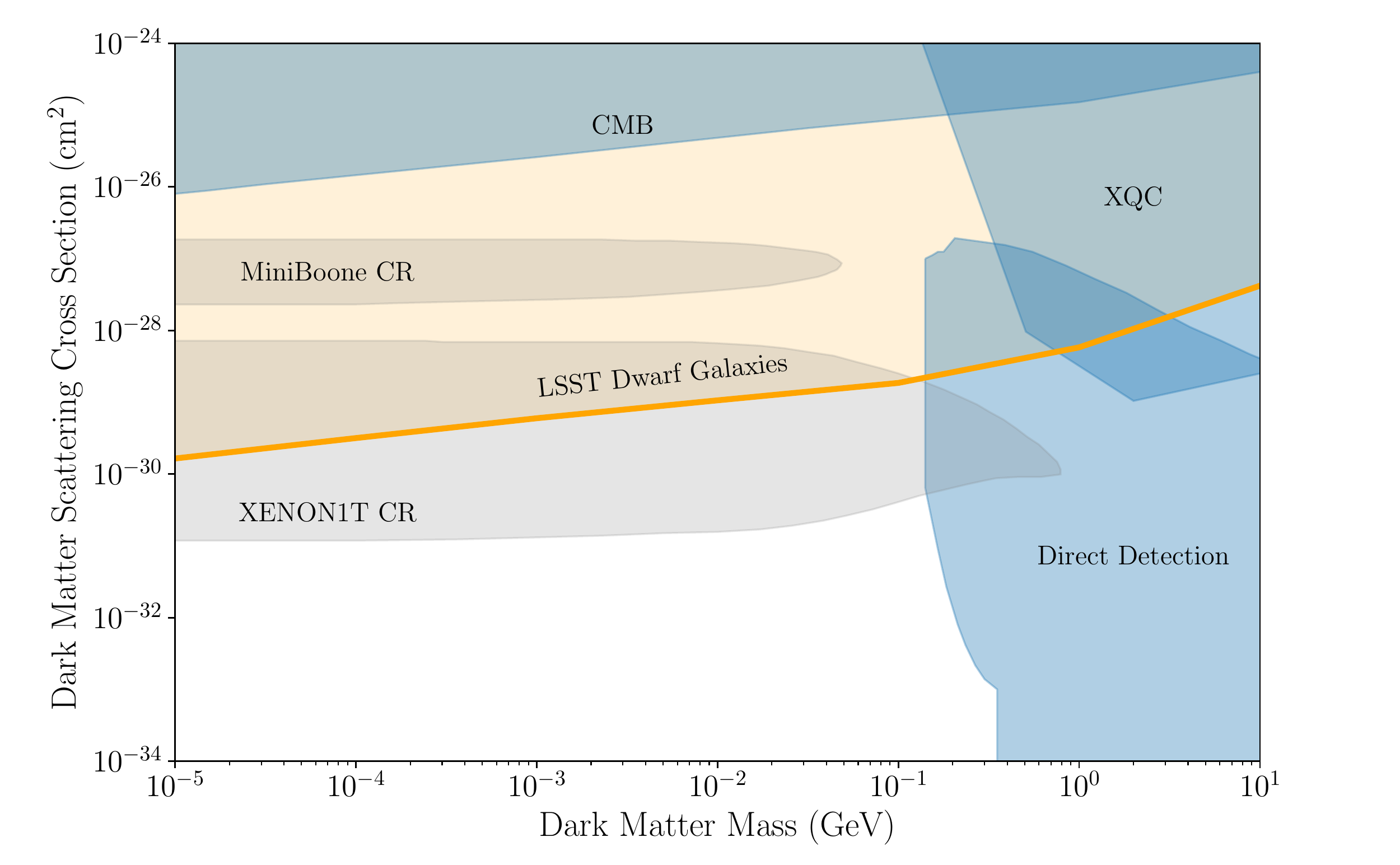}
\caption{\label{fig:macho_constraints}
    \emph{Left}: Constraints on the maximal fraction of dark matter in compact objects from existing probes (blue and gray) and projected sensitivity for LSST microlensing measurements (gold).
    \emph{Right}: Constraints on dark matter-baryon scattering through a velocity-independent, spin-independent contact interaction with protons from existing constraints (blue and gray) and projections for LSST observations of Milky Way satellite galaxies (gold) \citep{drlica-wagner_2019_lsst_dark_matter,Gluscevic:prep}.
    %Existing constraints are shown in blue and gray.
    %Existing constraints (shown in blue) include measurements of the CMB power spectrum \citep[CMB;][]{Gluscevic:2017ywp} and constraints from the X-ray Quantum Calorimeter experiment \citep[XQC;][]{0704.0794}. Direct detection constraints include results from CRESST-III \citep{1711.07692}, the CRESST 2017 surface run \citep{1707.06749}, and XENON1T \citep{1705.06655}, as interpreted by \citet[][]{1802.04764}. %\citep{2018PhRvD..97l3013K}.
    %Additional constraints that include the effects of cosmic-ray heating of dark matter are shown in gray \citep[][]{1810.10543}.
    %The projected sensitivity of LSST to dark matter-baryon scattering through observations of Milky Way satellite dwarf galaxies is shown in gold.
}
\end{figure}

\vspace{-1em} \subsection*{Recommendations for Astro 2020} \vspace{-0.5em}

LSST is scheduled to begin a decade of science operation in 2022; however, dark matter research with LSST is not yet funded.
Recognizing new opportunities created by LSST to constrain a range of dark matter models, we make the following recommendations to facilitate this science case:
%to facilitate the dark matter science goals outlined here:
%and in Drlica-Wagner et al. (2019)\citep{drlica-wagner_2019_lsst_dark_matter}:

\setlist{nolistsep}
\begin{itemize}[noitemsep]
%\begin{itemize}

    % (Funding agencies should)
    \item Support individual PIs and collaborative teams to analyze LSST data for dark matter science. %, leveraging the statistical power of LSST, and developing new techniques in image analysis, time-domain analysis, object classification, etc.
    %, to enhance sensitivity
    %(Funding agencies should)
    \item Support associated theoretical research to better understand the galaxy-halo connection, examine confounding baryonic effects, perform joint analyses of cosmological probes, investigate novel signatures of dark matter microphysics, and strengthen ties with the particle physics community. 
    %This work will likely involve dedicated numerical simulation activities.
    
    % (Funding agencies should)
    \item Support complementary observational facilities to investigate dark matter, including spectroscopic follow-up and high-resolution imaging, as well as multiwavelength analyses.
    %, as well as theoretical activities including joint-analysis of cosmological datasets and numerical simulations.
    %, as well as coordination of these efforts.

    % assemble into a cohesive community
    % The dark matter community should strive to present results in a manner that encourages... 
    %\item (Scientists) The astrophysical dark matter community should converge around a common language, and build  stronger connections with particle physics efforts.
    
    %\item (Scientists) 
    % Recognizing opportunities to enhance sensitivity and mitigate systematic uncertainty through the use fo joint probes,
    %The astrophysical dark matter community should strive to present results in ways that facilitate comparison between multiple probes and between theory and experiment.
    %The astrophysics and particle physics dark matter community should strive to present results in ways that facilitate comparison between multiple astrophysical probes and between theory and experiment.
    
    %\item (Scientists) The LSST dark matter science community should maintain/expand opportunities for collaboration. (This could mean finding a home within an existing Science Collaboration, forming a new collaboration, and/or establish strong connections between science collaborations. A key point is consistency between probes.)
    
    %\item (Scientists) Complementarity of astrophysical probes with particle physics experiment and theory. 
    
\end{itemize}

\noindent We anticipate that the multi-faceted LSST data will allow further probes of dark matter physics that have yet to be considered.
New ideas are especially important as searches for the most popular dark matter candidates gain in sensitivity while lacking a positive detection.
%the absence of evidence for the most popular dark matter candidates continues to grow.
As the particle physics community seeks to diversify the experimental effort to search for dark matter, it is important to remember that astrophysical observations provide robust, empirical measurement of fundamental dark matter properties.
In the coming decade, astrophysical observations will guide other experimental efforts, while simultaneously probing unique regions of dark matter parameter space.

\clearpage

\def\bibname{References}
\begingroup
  \small
  \setlength{\bibsep}{0pt plus 0.5ex}
  \bibliographystyle{JHEP}
  \bibliography{main}
\endgroup

\clearpage
\subsection*{Affiliations}

%%%%%%%%%%%%%%%%%%%%%%%%%%%%%%%%%%%%%%%%%%%%%%%%%%%%%%%%%%%%
%                                                          %
% Institutional aliases file.                              %
% Originally used in CV DM whitepaper,                     %
% successfully stolen for CV 21cm roadmap whitepaper,      %
% now to live a life of its own.                           %
%                                                          %
% When editing, please respect alphabetical ordering       %
% and avoid duplication (that is the entire point).        %
%                                                          %
%                                                          %
% A summer river being crossed                             %
% how pleasing                                             %
% with sandals in my hands!                                %
%       Yosa Buson (1716-1784)                             %
%                                                          %
%%%%%%%%%%%%%%%%%%%%%%%%%%%%%%%%%%%%%%%%%%%%%%%%%%%%%%%%%%%%

\newcommand{\Amherst}{University of Massachusetts, Amherst, MA 01003 USA}
\newcommand{\ANLHEP}{HEP Division, Argonne National Laboratory, Lemont, IL 60439, USA}
\newcommand{\APC}{Laboratoire Astroparticule et Cosmologie (APC), CNRS/IN2P3, Universit\'e Paris Diderot, 10, rue Alice Domon et Léonie Duquet, 75205 Paris Cedex 13, France}
\newcommand{\ASU}{Arizona State University, Tempe, AZ  85287}
\newcommand{\BenGurion}{Department of Physics, Ben-Gurion University, Be'er Sheva 84105, Israel}
\newcommand{\BNL}{Brookhaven National Laboratory, Upton, NY 11973}
\newcommand{\Brown}{Brown University, Providence, RI 02912}
\newcommand{\Bub}{Boston University, Boston, MA 02215}
\newcommand{\BU}{Boston University, Boston, MA 02215}
\newcommand{\Buffalo}{Department of Physics, University at Buffalo, SUNY Buffalo, NY 14260 USA}
\newcommand{\Caltech}{California Institute of Technology, Pasadena, CA 91125}
\newcommand{\Cardiff}{School of Physics and Astronomy, Cardiff University, The Parade, Cardiff, CF24 3AA, UK}
\newcommand{\Carleton}{Carleton University, K1S 5B6 Ottawa, Canada}
\newcommand{\Carnegie}{The Observatories of the Carnegie Institution for Science, 813 Santa Barbara St., Pasadena, CA 91101, USA}
\newcommand{\Cavendish}{Astrophysics Group, Cavendish Laboratory, J.J.Thomson Avenue, Cambridge, CB3 0HE, UK}
\newcommand{\CCA}{Center for Computational Astrophysics, 162 5th Ave, 10010, New York, NY, USA}
\newcommand{\CPPM}{Aix Marseille Univ, CNRS/IN2P3, CPPM, Marseille, France}
\newcommand{\CEADAP}{D\'epartement d’Astrophysique, CEA Saclay DSM/Irfu, 91191 Gif-sur-Yvette, France}
\newcommand{\CERN}{CERN, Geneva, Switzerland}
\newcommand{\CfA}{Harvard-Smithsonian Center for Astrophysics, MA 02138}
\newcommand{\CFT}{Center for Theoretical Physics, Polish Academy of Sciences, al. Lotnik\'{o}w 32/46, 02-668, Warsaw, Poland}
\newcommand{\Cincinnati}{University of Cincinnati, Cincinnati, OH 45221}
\newcommand{\CITA}{Canadian Institute for Theoretical Astrophysics, University of Toronto, Toronto, ON M5S 3H8, Canada}
\newcommand{\CNRSA}{CNRS, Laboratoire d'Annecy-le-Vieux de Physique Th\'{e}orique, France}
\newcommand{\CNYang}{C.N. Yang Institute for Theoretical Physics State University of New York Stony Brook, NY 11794}
\newcommand{\CMUCosmo}{Department 
of Physics, McWilliams Center for Cosmology, Carnegie Mellon University}
\newcommand{\Columbia}{Columbia University, New York, NY 10027}
\newcommand{\Cornell}{Cornell University, Ithaca, NY 14853}
\newcommand{\CPthree}{CP3-Origins, 5230 Odense, Denmark}
\newcommand{\CWRU}{Case Western Reserve University, Cleveland, OH 44106}
\newcommand{\daa}{Department of Astronomy and Astrophysics, University of Toronto, ON, M5S3H4}
\newcommand{\damtp}{DAMTP, Centre for Mathematical Sciences, Wilberforce Road, Cambridge, UK, CB3 0WA}
\newcommand{\DESY}{DESY,  22607 Hamburg, Germany}
\newcommand{\DFI}{Departamento de F\'isica, FCFM, Universidad de Chile, Blanco Encalada 2008, Santiago, Chile}
\newcommand{\DOE}{US. Department of Energy, Germantown, MD 20874}
\newcommand{\drexel}{Drexel University, Philadelphia, PA 19104}
\newcommand{\Duke}{Duke University and Triangle Universitites Nuclear Laboratory, Durham, NC 27708}
\newcommand{\DukePhys}{Department of Physics, Duke University, Durham, NC 27708, USA}
\newcommand{\dunlap}{Dunlap Institute for Astronomy and Astrophysics, University of Toronto, ON, M5S3H4}
\newcommand{\Durham}{Department of Physics, Lower Mountjoy, South Rd, Durham DH1 3LE, United Kingdom}
\newcommand{\ED}{University of Edinburgh, EH8 9YL Edinburgh, United Kingdom}
\newcommand{\EPFL}{Institute of Physics, Laboratory of Astrophysics, Ecole Polytechnique Fédérale de Lausanne (EPFL), Observatoire de Sauverny, 1290 Versoix, Switzerland}
\newcommand{\ETH}{ETH Zurich, Institute for Particle Physics, 8093 Zurich, Switzerland}
\newcommand{\FNAL}{Fermi National Accelerator Laboratory, Batavia, IL 60510}
\newcommand{\FQAUB}{Dept. de F\' isica Qu\` antica i Astrof\' isica, Universitat de Barcelona, Mart\' i i Franqu\` es 1, E08028 Barcelona, Spain}
\newcommand{\FSU}{Florida State University, Tallahassee, FL 32306}
\newcommand{\Glasgow}{University of Glasgow, G12 8QQ Glasgow, United Kingdom}
\newcommand{\GRAPPA}{GRAPPA Institute, University of Amsterdam, Science Park 904, 1098 XH Amsterdam, The Netherlands}
\newcommand{\GSFC}{Goddard Space Flight Center, Greenbelt, MD 20771 USA}
\newcommand{\GWU}{George Washington University, Washington, DC 20052}
\newcommand{\Hampton}{Hampton University, Hampton, VA 23668}
\newcommand{\HarvardPhys}{Department of Physics, Harvard University, Cambridge, MA 02138, USA}
\newcommand{\Haverford}{Haverford College, 370 Lancaster Ave, Haverford PA, 19041, USA}
\newcommand{\Hawaii}{University of Hawaii, Honolulu, HI 96822}
\newcommand{\HKUST}{The Hong Kong University of Science and Technology, Hong Kong SAR, China}
\newcommand{\houston}{University of Houston, Houston, TX 77204}
\newcommand{\IAP}{Institut d'Astrophysique de Paris (IAP), CNRS \& Sorbonne University, Paris, France}
\newcommand{\IAS}{Institute for Advanced Study, Princeton, NJ 08540}
\newcommand{\IBS}{Institute for Basic Science (IBS), Daejeon 34051, Korea}
\newcommand{\ICC}{ICC, University of Barcelona, IEEC-UB, Mart\' i i Franqu\` es, 1, E08028 Barcelona, Spain}
\newcommand{\ICCD}{Institute for Computational Cosmology, Department of Physics, Durham University, South Road, Durham, DH1 3LE, UK}
\newcommand{\ICE}{Institute of Space Sciences (ICE, CSIC), Campus UAB, Carrer de Can Magrans, s/n, 08193 Barcelona, Spain}
\newcommand{\ICRR}{Institute for Cosmic Ray Resaerch, The University of Tokyo, 456 Higashi-Mozumi, Kamioka, Hida, Gifu 506-1205, Japan}
\newcommand{\ICTP}{International Centre for Theoretical Physics, Strada Costiera, 11, I-34151 Trieste, Italy}
\newcommand{\IFAE}{Institut de Fisica d’Altes Energies, The Barcelona Institute of Science and Technology, Campus UAB, 08193 Bellaterra (Barcelona), Spain}
\newcommand{\IFPU}{IFPU - Institute for Fundamental Physics of the Universe, Via Beirut 2, 34014 Trieste, Italy}
\newcommand{\IFT}{Instituto de Fisica Teorica UAM/CSIC, Universidad Autonoma de Madrid, 28049 Madrid, Spain}
\newcommand{\IFUNAM}{IFUNAM - Instituto de F\'{i}sica, Universidad Nacional Aut\'onoma de M\'etico, 04510 CDMX, M\'exico}
\newcommand{\IHEP}{Institute of High Energy Physics, Austrian Academy of Sciences, 1050 Vienna, Austria}
\newcommand{\Imperial}{Theoretical Physics, Blackett Laboratory, Imperial College, London, SW7 2AZ, U.K.}
\newcommand{\Indiana}{Indiana University, Bloomington, IN 47405}
\newcommand{\INAFOATs}{INAF - Osservatorio Astronomico di Trieste, Via G.B. Tiepolo 11, 34143 Trieste, Italy}
\newcommand{\INAFOAS}{INAF - Osservatorio di Astrofisica e Scienza dello Spazio di Bologna, via Piero Gobetti 93/3, I-40129 Bologna, Italy}
\newcommand{\INFNCag}{Istituto Nazionale di Fisica Nucleare, Sezione di Cagliari,  09126 Cagliari, Italy}
\newcommand{\INFNCat}{Istituto Nazionale di Fisica Nucleare, Sezione di Catania, 95125 Catania, Italy}
\newcommand{\INFNG}{Istituto Nazionale di Fisica Nucleare, Sezione di Genova, 16146 Genova, Italy}
\newcommand{\INFN}{INFN – National Institute for Nuclear Physics, Via Valerio 2, I-34127 Trieste, Italy}
\newcommand{\INFNFE}{Istituto Nazionale di Fisica Nucleare, Sezione di Ferrara, 40122, Italy }
\newcommand{\INFNLNF}{Istituto Nazionale di Fisica Nucleare, Laboratori Nazionali di Frascati, 00044 Frascati, Italy}
\newcommand{\INFNLNS}{Istituto Nazionale di Fisica Nucleare, Laboratori Nazionali del Sud, 95125 Catania, Italy}
\newcommand{\INFNN}{Istituto Nazionale di Fisica Nucleare, Sezione di Napoli, 80125 Napoli, Italy }
\newcommand{\INFNRM}{Istituto Nazionale di Fisica Nucleare, Sezione di Roma, 00185 Roma, Italy}
\newcommand{\INFNT}{Istituto Nazionale di Fisica Nucleare, Sezione di Torino, 10125, Italy }
\newcommand{\ioa}{Institute of Astronomy, University of Cambridge,Cambridge CB3 0HA, UK}
\newcommand{\IPP}{Institute for Particle Physics, BC V8W 3P6 Victoria, Canada}
\newcommand{\IPMU}{Kavli Insitute for the Physics and Mathematics of the Universe (WPI), University of Tokyo, 277-8583 Kashiwa , Japan}
\newcommand{\IPNL}{Universit\'e de Lyon, F-69622, Lyon, France; Universit\'e de Lyon 1, Villeurbanne; CNRS/IN2P3, Institut de Physique Nucl\'eaire de Lyon}
\newcommand{\IRFU}{IRFU, CEA, Universit\'e Paris-Saclay, F-91191 Gif-sur-Yvette, France}
\newcommand{\ITFA}{Institute for Theoretical Physics, University of Amsterdam, Science Park 904, 1098 XH Amsterdam, The Netherlands}
\newcommand{\IUCAA}{The Inter-University Centre for Astronomy and Astrophysics, Pune, 411007, India}
\newcommand{\Jerusalem}{Hebrew University of Jerusalem, 91904 Jerusalem, Israel}
\newcommand{\JHU}{Johns Hopkins University, Baltimore, MD 21218}
\newcommand{\JLAB}{Thomas Jefferson National Laboratory, Newport News, VA 23606}
\newcommand{\JPL}{Jet Propulsion Laboratory, California Institute of Technology, Pasadena, CA, USA}
\newcommand{\KASSI}{Korea Astronomy and Space Science Institute, Daejeon 34055, Korea}
\newcommand{\kavli}{Kavli Institute for Cosmology, Cambridge, UK, CB3 0HA}
\newcommand{\KIAS}{School of Physics, Korea Institute for Advanced Study, 85 Hoegiro, Dongdaemun-gu, Seoul 130-722, Korea}
\newcommand{\KICP}{Kavli Institute for Cosmological Physics, Chicago, IL 60637}
\newcommand{\KIPAC}{Kavli Institute for Particle Astrophysics and Cosmology, Stanford 94305}
\newcommand{\KINGS}{King's College London, WC2R 2LS London, United Kingdom}
\newcommand{\Kobe}{Kobe University, 657-8501 Kobe, Japan}
\newcommand{\KPH}{Johannes Gutenberg University, 55128 Mainz, Germany}
\newcommand{\KPMU}{University of Tokyo, 277-8583  Kashiwa , Japan}
\newcommand{\KSU}{Kansas State University, Manhattan, KS 66506}
\newcommand{\Lafayette}{Lafayette College, Easton, PA 18042}
\newcommand{\LANL}{Los Alamos National Laboratory, Los Alamos, NM 87545}
\newcommand{\LBL}{Lawrence Berkeley National Laboratory, Berkeley, CA 94720}
\newcommand{\Leiden}{Lorentz Institute, Leiden University, Niels Bohrweg 2,Leiden, NL 2333 CA, The Netherlands}
\newcommand{\Liverpool}{University of Liverpool,  L69 7ZE Liverpool , United Kingdom}
\newcommand{\LLNL}{Lawrence Livermore National Laboratory, Livermore, CA, 94550}
\newcommand{\LPC}{Universit\'e Clermont Auvergne, CNRS/IN2P3, Laboratoire de Physique de Clermont, F-63000 Clermont-Ferrand, France}
\newcommand{\LPNHE}{Sorbonne Universit\'e, Universit\'e Paris Diderot, CNRS/IN2P3, Laboratoire de Physique Nucl\'eaire et de Hautes Energies, LPNHE, 4 Place Jussieu, F-75252 Paris, France}
\newcommand{\McGill}{McGill University, Montreal, QC H3A 2T8, Canada}
\newcommand{\Melbourne}{School of Physics, The University of Melbourne, Parkville, VIC 3010, Australia}
\newcommand{\Mines}{Colorado School of Mines, Golden, CO 80401}
\newcommand{\MIT}{Massachusetts Institute of Technology, Cambridge, MA 02139}
\newcommand{\MPE}{Max-Planck-Institut f\"{u}r extraterrestrische Physik (MPE), Giessenbachstrasse 1, D-85748 Garching bei M\"unchen, Germany}
\newcommand{\MPIA}{Max-Planck-Institut f\"{u}r Astrophysik, Karl-Schwarzschild-Str. 1, 85741 Garching, Germany}
\newcommand{\MPP}{Max-Planck-Institut f\"{u}r Physik (Werner-Heisenberg-Institut), F\"ohringer Ring 6, D-80805 M\"unchen, Germany}
\newcommand{\LUPM}{Laboratoire Univers et Particules de Montpellier, Univ. Montpellier and CNRS, 34090 Montpellier, France}
\newcommand{\NAOC}{National Astronomical Observatories, Chinese Academy of Sciences, PR China}
\newcommand{\NCBJ}{National Center for Nuclear Research, Ul.Pasteura 7,Warsaw, Poland}
\newcommand{\NCU}{National Central University, Taoyuan City 32001, Taiwan (R.O.C.)}
\newcommand{\NCSU}{Physics Department, North Carolina State Universitym, 2401 Stinson Dr, Raleigh, NC 27607}
\newcommand{\ND}{University of Notre Dame,vNotre Dame, IN 46556}
\newcommand{\NIU}{Northern Illinois University, DeKalb, Illinois 60115}
\newcommand{\NMSU}{New Mexico State University, Las Cruces, NM 88003}
\newcommand{\NOAO}{National Optical Astronomy Observatory, 950 N. Cherry Ave., Tucson, AZ 85719 USA}
\newcommand{\Northwestern}{Northwestern University, Evanston, IL 60201}
\newcommand{\Nottingham}{University of Nottingham, NG7 2RD Nottingham, United Kingdom}
\newcommand{\NWU}{Northwestern University, Evanston, IL 60208}
\newcommand{\NYU}{New York University, New York, NY 10003}
\newcommand{\OK}{ University of Oklahoma, Norman, OK 73019}
\newcommand{\ORNL}{Oak Ridge National Laboratory, Oak Ridge, TN 37831}
\newcommand{\OSU}{The Ohio State University, Columbus, OH 43212}
\newcommand{\OU}{Department of Physics and Astronomy, Ohio University, Clippinger Labs, Athens, OH 45701, USA}
\newcommand{\OskarKlein}{Oskar Klein Centre for Cosmoparticle Physics, Stockholm University, AlbaNova, Stockholm SE-106 91, Sweden}
\newcommand{\Oxford}{The University of Oxford, Oxford OX1 3RH, UK}
\newcommand{\Oxy}{Occidental College, Los Angeles, CA 90041}
\newcommand{\ParisSud}{Universit\'{e} Paris-Sud, LAL, UMR 8607, F-91898 Orsay Cedex, France \& CNRS/IN2P3, F-91405 Orsay, France}
\newcommand{\PI}{Perimeter Institute, Waterloo, Ontario N2L 2Y5, Canada}
\newcommand{\Pitt}{University of Pittsburgh and PITT PACC, Pittsburgh, PA 15260}
\newcommand{\PNNL}{Pacific Northwest National Laboratory ,Richland, WA 99352}
\newcommand{\PNPI}{Petersburg Nuclear Physics Institute, 188300 Gatchina, Russia}
\newcommand{\Port}{Institute of Cosmology \& Gravitation, University of Portsmouth, Dennis Sciama Building, Burnaby Road, Portsmouth PO1 3FX, UK}
\newcommand{\Princeton}{Princeton University, Princeton, NJ 08544}
\newcommand{\PSU}{The Pennsylvania State University, University Park, PA 16802}
\newcommand{\Purdue}{Purdue University, West Lafayette, IN 47907}
\newcommand{\PW}{Participation Worldscope, Sedona, Arizona and Hong Kong, SAR PRC}
\newcommand{\Queens}{Queen's University , K7L 3N6 Kingston, Canada}
\newcommand{\Queensland}{The University of Queensland, School of Mathematics and Physics, QLD 4072, Australia}
\newcommand{\QMUL}{Queen Mary University of London, Mile End Road, London E1 4NS, United Kingdom}
\newcommand{\RAL}{Radio Astronomy Laboratory, University of California Berkeley, Berkeley, CA 94720, USA}
\newcommand{\Rice}{Department of Physics \& Astronomy, Rice University, Houston, Texas 77005, USA}
\newcommand{\RIT}{Rochester Institute of Technology}
\newcommand{\RomaS}{Dipartimento di Fisica, Universit\`{a} La Sapienza, P. le A. Moro 2, Roma, Italy}
\newcommand{\RUG}{Kapteyn Astronomical Institute, University of Groningen, P.O. Box 800, 9700 AV Groningen, The Netherlands}
\newcommand{\Rutgers}{Department of Physics and Astronomy, Rutgers, the State University of New Jersey, 136 Frelinghuysen Road, Piscataway, NJ 08854, USA}
\newcommand{\Sanford}{Sanford Underground Research Facility, Lead, SD 57754}
\newcommand{\Sassari}{Universit\`a di Sassari, 07100 Sassari,  Italy}
\newcommand{\SCIPP}{University of California at Santa Cruz, Santa Cruz, CA 95064}
\newcommand{\Sejong}{Department of Physics and Astronomy, Sejong University, Seoul, 143-747, Korea}
\newcommand{\Sheffield}{University of Sheffield, S3 7RH Sheffield, United Kingdom}
\newcommand{\SHAO}{Shanghai Astronomical Observatory (SHAO), Nandan Road 80, Shanghai 200030, China}
\newcommand{\Siena}{Siena College, 515 Loudon Road, Loudonville, NY 12211, USA}
\newcommand{\SISSA}{SISSA - International School for Advanced Studies, Via Bonomea 265, 34136 Trieste, Italy}
\newcommand{\SimonFraser}{Department of Physics, Simon Fraser University, Burnaby, British Columbia, Canada V5A 1S6}
\newcommand{\SLAC}{SLAC National Accelerator Laboratory, Menlo Park, CA 94025}
\newcommand{\SMU}{Southern Methodist University, Dallas, TX 75275}
\newcommand{\SNOLAB}{SNOLAB, Lively, ON P3Y 1N2, Canada}
\newcommand{\SoCal}{University of Southern California, CA 90089 }
\newcommand{\Stanford}{Stanford University, Stanford, CA 94305}
\newcommand{\StonyBrook}{Stony Brook University, Stony Brook, NY 11794}
\newcommand{\STSCI}{Space Telescope Science Institute, Baltimore, MD 21218}
\newcommand{\SUNYA}{University at Albany SUNY, Albany, NY 12222}
\newcommand{\SussexAstronomy}{Astronomy Centre, School of Mathematical and Physical Sciences, University of Sussex, Brighton BN1 9QH, United Kingdom}
\newcommand{\Syracuse}{Syracuse University, Syracuse, NY 13244}
\newcommand{\Tamu}{Texas AandM University, College Station, TX 77843 }
\newcommand{\Techsource}{Techsource Incorporated, Los Alamos, NM 87544}
\newcommand{\TelAviv}{Tel-Aviv University,  69978 Tel-Aviv, Israel}
\newcommand{\Temple}{Temple University, Philadelphia, PA 19122}
\newcommand{\TIFR}{Tata Institute of Fundamental Research, Homi Bhabha Road, Mumbai 400005 India}
\newcommand{\Tsinghua}{Department of Physics and Tsinghua Center for Astrophysics, Tsinghua University, Beijing 100084, P R China}
\newcommand{\TUM}{Technical University of Munich,  80333 Munich, Germany}
\newcommand{\UA}{University of Alabama, Tuscaloosa, AL 35487}
\newcommand{\UAS}{Department of Astronomy/Steward Observatory, University of Arizona, Tucson, AZ  85721}
\newcommand{\UAM}{Universidad Aut\'onoma de Madrid, 28049, Madrid, Spain}
\newcommand{\UBC}{University of British Columbia, Vancouver, BC V6T 1Z1, Canada}
\newcommand{\UCB}{Department of Astronomy, University of California Berkeley, Berkeley, CA 94720, USA}
\newcommand{\UCBP}{Department of Physics, University of California Berkeley, Berkeley, CA 94720, USA}
\newcommand{\UCBSSL}{Space Sciences Laboratory, University of California Berkeley, Berkeley, CA 94720, USA}
\newcommand{\UCD}{University of California at Davis, Davis, CA 95616}
\newcommand{\UChicago}{University of Chicago, Chicago, IL 60637}
\newcommand{\UCI}{University of California, Irvine, CA 92697}
\newcommand{\UCLA}{University of California at Los Angeles, Los Angeles,  CA 90095}
\newcommand{\UCL}{University College London, WC1E 6BT London, United Kingdom}
\newcommand{\UCR}{University of California at Riverside, Riverside, CA 92521}
\newcommand{\UCSB}{University of California at Santa Barbara, Santa Barbara, CA 93106}
\newcommand{\UCSC}{University of California at Santa Cruz, Santa Cruz, CA 95064}
\newcommand{\UCSD}{University of California San Diego, La Jolla, CA 92093}
\newcommand{\UFL}{University of Florida, Gainesville, FL 32611}
\newcommand{\UFN}{Universit\`a Federico II di Napoli, 80125 Napoli, Italy}
\newcommand{\UGTO}{Divisi\'on de Ciencias e Ingenier\'ias, Universidad de Guanajuato, Le\'on 37150, M\'exico}
\newcommand{\UKY}{University of Kentucky, Lexington, KY 40506}
\newcommand{\UMD}{University of Maryland, College Park, MD 20742
\newcommand{\UMiami}{University of Miami, Coral Gables, FL 33124}}
\newcommand{\UMich}{University of Michigan, Ann Arbor, MI 48109}
\newcommand{\UMN}{University of Minnesota, Minneapolis, MN 55455}
\newcommand{\UnB}{Instituto de F\'{i}sica, Universidade de Bras\'{i}lia, 70919-970, Bras\'{i}lia, DF, Brazil}
\newcommand{\UNC}{University of North Carolina at Chapel Hill, Chapel Hill, NC 27599}
\newcommand{\UNH}{University of New Hampshire, Durham, NH 03824}
\newcommand{\UNIMI}{Dipartimento di Fisica ``Aldo Pontremoli'', Universit\`a{} degli Studi di Milano, via Celoria 16, 20133 Milano, Italy}
\newcommand{\UNIPD}{Dipartimento di Fisica e Astronomia ``G. Galilei'',Universit\`a degli Studi di Padova, via Marzolo 8, I-35131, Padova, Italy}
\newcommand{\UNM}{University of New Mexico, Albuquerque, NM 87131}
\newcommand{\UNV}{University of Nevada, Reno, NV 89557}
\newcommand{\UoM}{Jodrell Bank Center for Astrophysics, School of Physics and Astronomy, University of Manchester, Oxford Road, Manchester, M13 9PL, UK}
\newcommand{\UPenn}{Department of Physics and Astronomy, University of Pennsylvania, Philadelphia, Pennsylvania 19104, USA}
\newcommand{\UR}{Department of Physics and Astronomy, University of Rochester, 500 Joseph C. Wilson Boulevard, Rochester, NY 14627, USA}
\newcommand{\UrbanaC}{Department of Physics, University of Illinois at Urbana-Champaign, Urbana, Illinois 61801, USA}
\newcommand{\USC}{The University of South Carolina, Columbia, SC 29208}
\newcommand{\USD}{The University of South Dakota, Vermillion, SD 57069}
\newcommand{\UTD}{University of Texas at Dallas, Texas 75080}
\newcommand{\Utenn}{The University of Tennessee, Knoxville, TN 37996}
\newcommand{\Utah}{University of Utah, Department of Physics and Astronomy, 115 S 1400 E, Salt Lake City, UT 84112, USA}
\newcommand{\UVA}{University of Virginia, Charlottesville, VA 22903}
\newcommand{\Uvic}{University of Victoria, BC V8P 5C2 Victoria, Canada}
\newcommand{\UWaterloo}{Department of Physics and Astronomy, University of Waterloo, 200 University Ave W, Waterloo, ON N2L 3G1, Canada}
\newcommand{\UWMadison}{Department of Physics, University of Wisconsin - Madison, Madison, WI 53706}
\newcommand{\UW}{University of Washington, Seattle 98195}
\newcommand{\UWC}{Department of Physics \& Astronomy, University of the Western Cape, Cape Town 7535, South Africa}
\newcommand{\Vanderbilt}{Physics \& Astronomy Department, Vanderbilt University, PMB 401807, 2301 Vanderbilt Place, Nashville, TN 37235}
\newcommand{\VSI}{Van Swinderen Institute for Particle Physics and Gravity, University of Groningen, Nijenborgh 4, 9747~AG~Groningen, The~Netherlands}
\newcommand{\VT}{Virginia Tech, Blacksburg, VA 24061}
\newcommand{\VUU}{Virginia Union University, Richmond, Virginia, 23220}
\newcommand{\WCA}{Centre for Astrophysics, University of Waterloo, Waterloo, Ontario N2L 3G1, Canada}
\newcommand{\Weizmann}{Weizmann Institute of Science, 76100 Rehovot, Israel}
\newcommand{\Wellesley}{Wellesley College, Wellesley, MA 02481}
\newcommand{\wiscIce}{University of Wisconsin, Madison, WI 53706}
\newcommand{\WM}{College of William and Mary, Newport News, VA 23606}
\newcommand{\WUSL}{Washington University in St Louis, St. Louis, MO 63130}
\newcommand{\WVU}{CSEE, West Virginia University, Morgantown, WV 26505, USA}
\newcommand{\WVUGWAC}{Center for Gravitational Waves and Cosmology, West Virginia University, Morgantown, WV 26505, USA}
\newcommand{\Wyoming}{Department of Physics and Astronomy, University of Wyoming, Laramie, WY 82071, USA}
\newcommand{\Yale}{Department of Physics, Yale University, New Haven, CT 06520}
\newcommand{\YorkU}{Department of Physics and Astronomy, York University, Toronto, Ontario M3J 1P3, Canada}

\begin{multicols}{2}
\scriptsize
\parskip=4pt

\affil{$^{1}$ \UWMadison}
\affil{$^{2}$ \FNAL}
\affil{$^{3}$ \KICP}
\affil{$^{4}$ \UChicago}
\affil{$^{5}$ \UCI}
\affil{$^{6}$ \damtp}
\affil{$^{7}$ \KIPAC}
\affil{$^{8}$ \NYU}
\affil{$^{9}$ \Durham}
\affil{$^{10}$ \LLNL}
\affil{$^{11}$ \KASSI}
\affil{$^{12}$ \SISSA}
\affil{$^{13}$ \IFPU}
\affil{$^{14}$ \INFN}
\affil{$^{15}$ \SLAC}
\affil{$^{16}$ \GRAPPA}
\affil{$^{17}$ \Leiden}
\affil{$^{18}$ \JHU}
\affil{$^{19}$ \Port}
\affil{$^{20}$ \UCR}
\affil{$^{21}$ \UCLA}
\affil{$^{22}$ \OskarKlein}
\affil{$^{23}$ \INAFOATs}
\affil{$^{24}$ \EPFL}
\affil{$^{25}$ \CfA}
\affil{$^{26}$ \LBL}
\affil{$^{27}$ \daa}
\affil{$^{28}$ \UAS}
\affil{$^{29}$ \Rutgers}
\affil{$^{30}$ \Stanford}
\affil{$^{31}$ \PI}
\affil{$^{32}$ \CNRSA}
\affil{$^{33}$ NASA Goddard Space Flight Center}
\affil{$^{34}$ \UCBP}
\affil{$^{35}$ \OU}
\affil{$^{36}$ \LUPM}
\affil{$^{37}$ \MPE}
\affil{$^{38}$ \CMUCosmo}
\affil{$^{39}$ Institute for Theoretical Particle Physics and Cosmology, RWTH Aachen University, Germany}
\affil{$^{40}$ Univ. Grenoble Alpes, USMB, CNRS, LAPTh, F-74940 Annecy, France}
\affil{$^{41}$ \HarvardPhys}
\affil{$^{42}$ \UNM}
\affil{$^{43}$ \Queensland}
\affil{$^{44}$ \IFUNAM'}
\affil{$^{45}$ \UoM}
\affil{$^{46}$ \JPL}
\affil{$^{47}$ Laboratory for Astroparticle Physics, University of Nova Gorica}
\affil{$^{48}$ Department of Physics, University of Surrey, UK}
\affil{$^{49}$ \UCD}
\affil{$^{50}$ \Oxford}
\affil{$^{51}$ \CITA}
\affil{$^{52}$ \kavli}
\affil{$^{53}$ \IFT}
\affil{$^{54}$ \ANLHEP}
\affil{$^{55}$ Physical Science Department, Barry University}
\affil{$^{56}$ \UFL}
\affil{$^{57}$ \UR}
\affil{$^{58}$ \UGTO}
\affil{$^{59}$ \Haverford}
\affil{$^{60}$ \CCA}
\affil{$^{61}$ \OSU}
\affil{$^{62}$ \dunlap}
\affil{$^{63}$ \VT}
\affil{$^{64}$ \UPenn}
\affil{$^{65}$ Yonsei University, Seoul, South Korea}
\affil{$^{66}$ \UCSC}
\affil{$^{67}$ \TIFR}
\affil{$^{68}$ \ioa}
\affil{$^{69}$ \Brown}
\affil{$^{70}$ \BenGurion}
\affil{$^{71}$ \UCL}
\affil{$^{72}$ \UCB}
\affil{$^{73}$ \NOAO}
\affil{$^{74}$ \UNIPD}
\affil{$^{75}$ \UCSD}
\affil{$^{76}$ \Princeton}
\affil{$^{77}$ \CNYang}
\affil{$^{78}$ \Pitt}
\affil{$^{79}$ \VSI}
\affil{$^{80}$ \ICTP}
\affil{$^{81}$ Laboratoire de l'Accélérateur Linéaire, IN2P3-CNRS, France}
\affil{$^{82}$ \IUCAA}
\affil{$^{83}$ \Siena}
\affil{$^{84}$ \Wyoming}
\affil{$^{85}$ \Caltech}
\affil{$^{86}$ \Yale}
\affil{$^{87}$ \BNL}
\affil{$^{88}$ \Tamu}
\affil{$^{89}$ \ETH}
\affil{$^{90}$ Center for Cosmology and AstroParticle Physics, The Ohio State University}
\affil{$^{91}$ Department of Astronomy, The Ohio State University}
\affil{$^{92}$ \RomaS}
\affil{$^{93}$ \INFNRM}
\affil{$^{94}$ \UNH}
\affil{$^{95}$ \STSCI}
\affil{$^{96}$ Laborat\'orio Interinstitucional de e-Astronomia - LIneA, Rua Gal. Jos\'e Cristino 77, Rio de Janeiro, RJ - 20921-400, Brazil}
\affil{$^{97}$ \Sejong}
\affil{$^{98}$ \KSU}
\affil{$^{99}$ Departamento de F\'isica Te\'orica, M-15, Universidad Aut\'onoma de Madrid, E-28049 Madrid, Spain}
\affil{$^{100}$ \StonyBrook}
\affil{$^{101}$ \SHAO}
\affil{$^{102}$ \Carnegie}
\affil{$^{103}$ \UMich}
\affil{$^{104}$ \MIT}
\affil{$^{105}$ INAF-Italian National Institute of Astrophysics, Italy}
\affil{$^{106}$ Space Telescope Science Institute}
\affil{$^{107}$ \DukePhys}
\affil{$^{108}$ \Duke}
\affil{$^{109}$ \UGTO'}
\affil{$^{110}$ \houston}
\affil{$^{111}$ \Syracuse}
\affil{$^{112}$ \MPIA}
\affil{$^{113}$ Center for Astrophysics and Cosmology, University of Nova Gorica}
\affil{$^{114}$ \ED}

\normalsize
\end{multicols}
\parskip=8pt

\end{document}